\documentclass[preprint,showpacs,preprintnumbers,amsmath,amssymb]{revtex4}
\newcommand{\newscale}{0.4}
\newcommand{\newscalex}{0.65}
\newcommand{\newscaley}{0.1}
\newcommand{\shiftc}{2.65cm}
\newcommand{\shiftb}{2.85cm}
\newcommand{\shifta}{2.75cm}

\usepackage{graphicx,verbatim,amssymb}
\usepackage{dcolumn}
\usepackage{bm}
\usepackage{sidecap}
\graphicspath{{/user/rhaertle/epsfiles/Propplots/},{/user/rhaertle/Berlin08/},
{/user/rhaertle/epsfiles/Rate-Report/},{/user/rhaertle/epsfiles/RateCoul/},
{/user/rhaertle/epsfiles/Benesch/},{/user/rhaertle/PaperI/},{/user/rhaertle/PaperII/},
{/user/rhaertle/epsfiles/Cizek/},
{/user/rhaertle/epsfiles/PaperIV/},
{/user/rhaertle/epsfiles/Oskar-Report/},{/user/rhaertle/PaperOskar/},{/user/rhaertle/PaperDeph/}}

\begin{document}

\title{Quantum Interference and Decoherence in Single-Molecule Junctions: 
How Vibrations Induce Electrical Current}

\author{R.\ H\"artle$^{1}$}
\author{M.\ Butzin$^{1}$}
\author{O.\ Rubio-Pons$^{1,2}$}
\author{M.\ Thoss$^{1}$}%
\affiliation{%
$^{1}$ Institut f\"ur Theoretische Physik und 
Interdisziplin\"ares Zentrum f\"ur Molekulare Materialien, 
Friedrich-Alexander-Universit\"at Erlangen-N\"urnberg, 
Staudtstr.\,7/B2, D-91058 Erlangen, Germany\\
$^{2}$ Theoretische Chemie, 
Technische Universit\"at M\"unchen, 
Lichtenbergstr.\,4, D-85747 Garching, Germany
}%

\date{\today}

\begin{abstract}
Quantum interference effects and 
decoherence mechanisms in single-molecule junctions are analyzed 
employing a nonequilibrium Green's function approach. 
Electrons tunneling through quasi-degenerate states of a nanoscale molecular 
junction exhibit interference effects. 
We show that  
electronic-vibrational coupling, inherent to any molecular junction, 
strongly quenches such interference effects. 
As a result, the electrical current can be significantly larger 
than without electronic-vibrational coupling. 
The analysis reveals that the quenching of quantum interference is particularly pronounced
if the junction is vibrationally highly excited, e.g.\ due to current-induced
nonequilibrium effects in the resonant transport regime.
\end{abstract}


\maketitle

Quantum interference 
is at the heart of quantum mechanics. 
While this classically counter-intuitive phenomenon has been verified long ago,
\emph{e.g.}\ by double-slit experiments \cite{Moellenstedt1956,Zeilinger}, 
quantum interference effects 
recently attracted much attention in electron transport through nanostructures, 
such as, \emph{e.g.}, quantum dots \cite{Holleitner2001,Kubala2002,Donarini2010,Markussen2010,Ueda2010} 
or single-molecule junctions 
\cite{Solomon2006,Brisker2008,Darau2009,Schultz2010}. 
When a molecule is contacted by two electrodes, 
forming a single-molecule junction \cite{Galperin07,cuevasscheer2010}, 
two different physical regimes become interconnected: 
The microscopic realm of a single molecule, governed by coherent quantum dynamics, 
and macroscopic reservoirs of electrons, 
wherein electronic coherences decay rapidly.  
Hence, the question arises to which extent a single-molecule 
junction preserves quantum coherence and which decoherence mechanisms 
are active. Understanding these mechanisms is 
crucial for the design of nanoelectronic devices. 

\begin{figure}
\includegraphics[scale=\newscale]{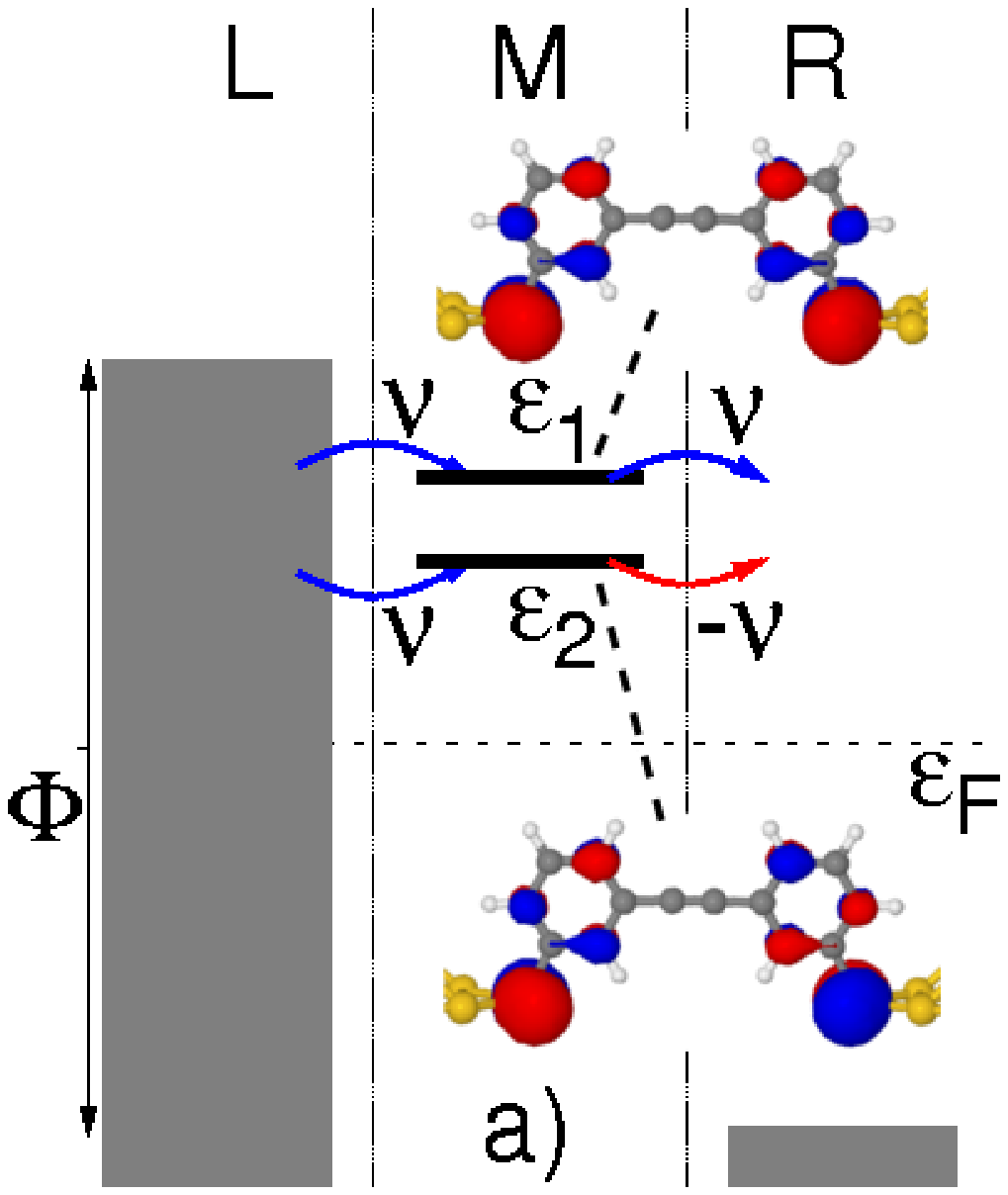}
\includegraphics[scale=\newscale]{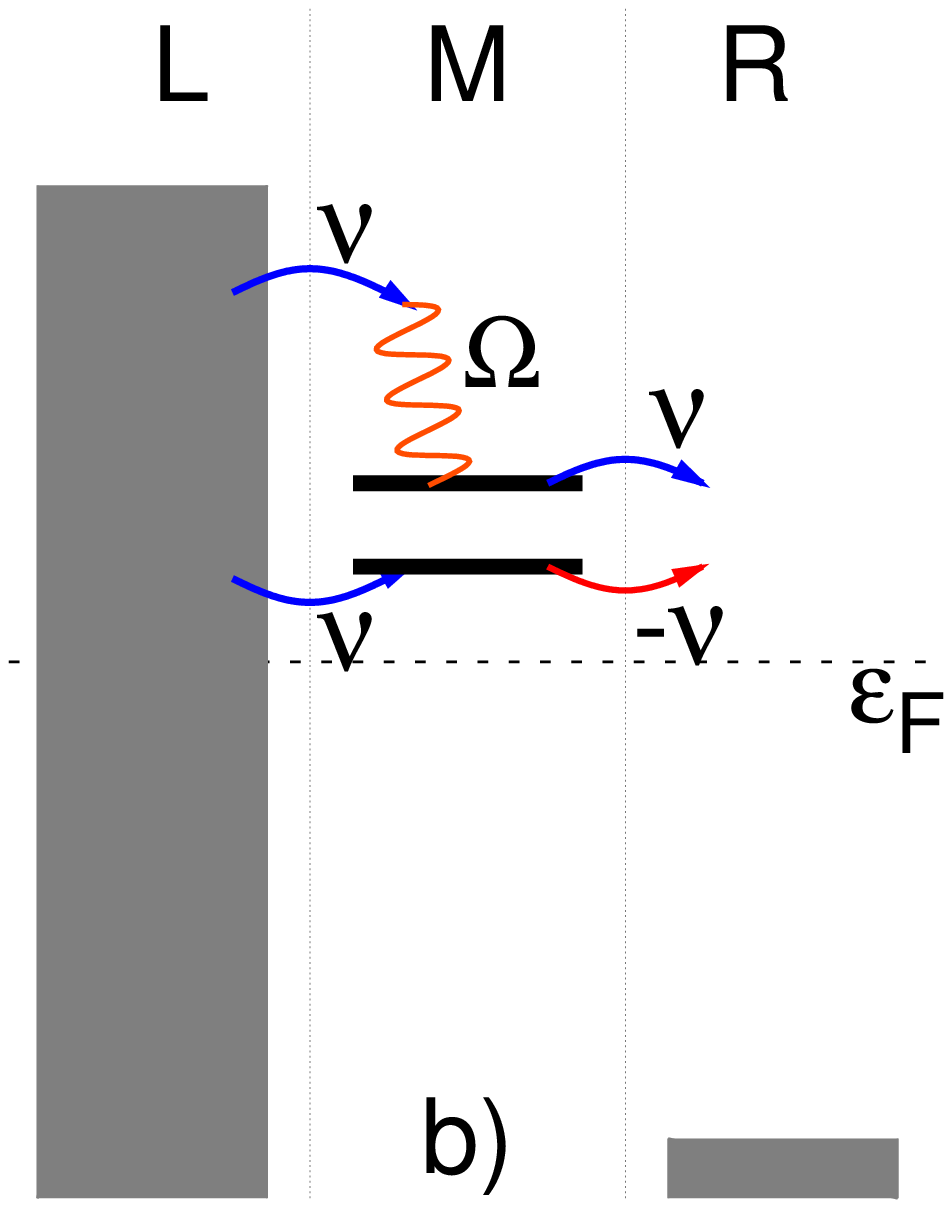}
\includegraphics[scale=\newscale]{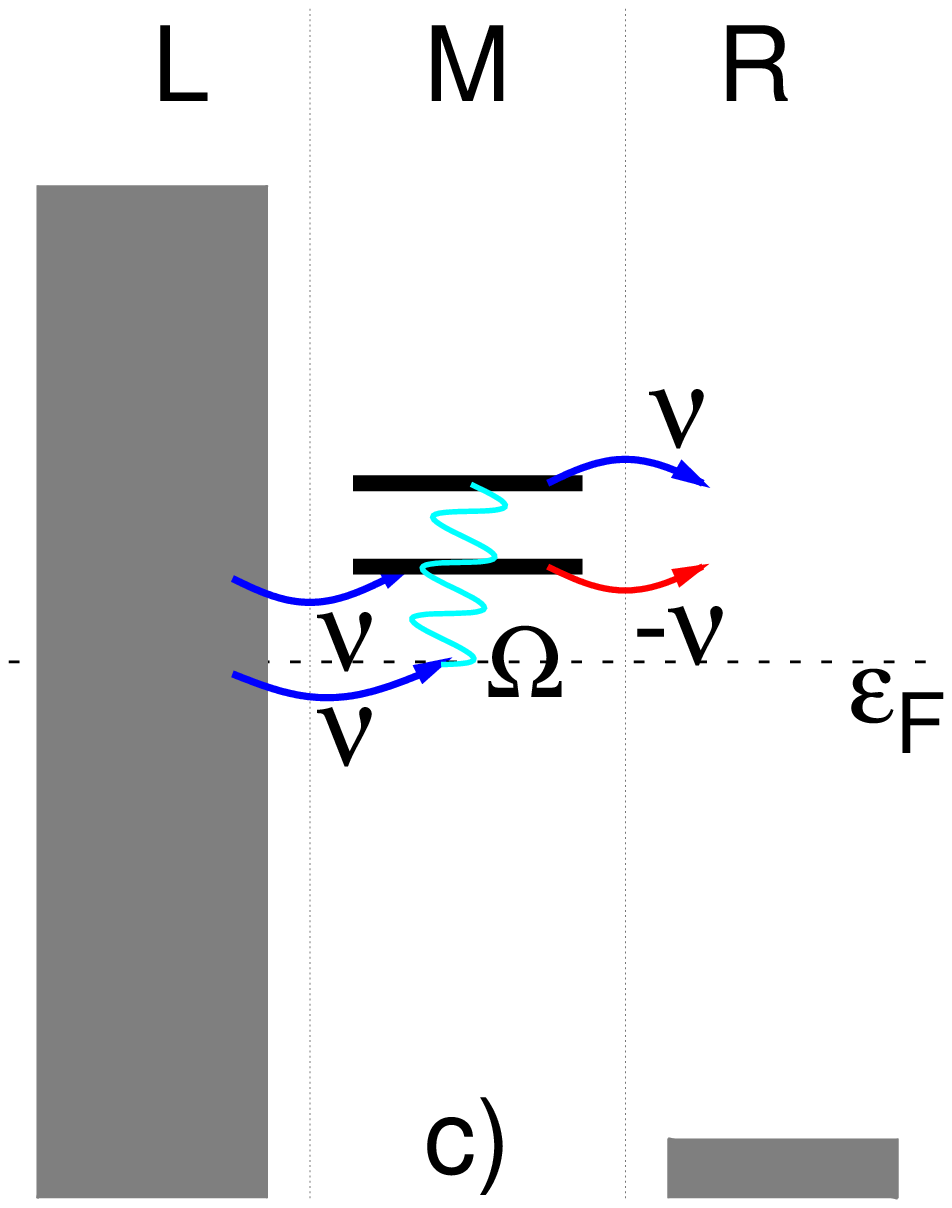}
\caption{(Color online) \label{scetchofmodelsystem} 
a): Scheme of a molecular junction 
at a finite bias voltage $\Phi$ that exhibits destructive interference. 
The gray shaded areas depict the continuum of occupied electronic states 
in the left (L) and the right (R) lead. Electronic states located at the 
molecular bridge (M) are represented by black bars. 
The blue and red arrows depict the coupling of these states to the leads, 
which due to the different symmetry of the corresponding orbitals, 
\emph{gerade} and \emph{ungerade}, differ by sign 
(example orbitals are given in the insets). 
b) and c): Example processes for vibrational excitation 
(red wiggly line) and deexcitation (blue wiggly line) 
upon inelastic tunneling of an electron  
through the molecular junction.}
\end{figure}

This question can be addressed by analyzing 
quantum interference effects that occur in a single-molecule junction. Consider
\emph{e.g.}\ the model molecular junction depicted in Fig.\ \ref{scetchofmodelsystem}a. 
This junction comprises two electronic states, 
which are located at the molecular bridge. One of them is 
\emph{gerade} with respect to the $\text{L}\leftrightarrow\text{R}$ symmetry
of the junction,  the other is \emph{ungerade}. 
 If these states are quasi-degenerate, they provide, 
similar to a double-slit experiment, two 
different pathways for an electron tunneling through the junction. 
As the respective wave-functions for the outgoing electron thus differ by sign, 
they destructively interfere with each other. As a result, 
the corresponding tunnel current is suppressed, or may even completely vanish
({\it vide infra}). 
In this letter we show that such quantum interference effects 
are quenched by vibrationally-assisted tunneling processes 
(cf.\ Fig.\ \ref{scetchofmodelsystem}b and \ref{scetchofmodelsystem}c), 
during which the tunneling electron excites or deexcites vibrational 
degrees of freedom of the junction. 
Due to the small size and mass of a molecular conductor, 
vibrational and electronic 
degrees of freedom are typically strongly coupled, resulting in 
inelastic tunneling processes, which strongly influence 
the respective transport characteristics 
\cite{Galperin07,Flensberg03,Mitra04,Lueffe,Hartle,Hartle09,Hartle2010,Hartle2010b}. 
This aspect distinguishes nanoscale molecular 
conductors from mesoscopic systems and quantum dots. 

We describe vibrationally coupled electron
transport through a single molecule (M) that is
covalently bound to two metal leads (L,R) by a generic model with Hamiltonian 
$H=H_{\text{el}}+H_{\text{vib}}$, where ($\hbar=1$)
\begin{eqnarray}
H_{\text{el}}\hspace{-1.5mm} &=& 
\hspace{-1.5mm}\sum_{i\in\text{M}} \epsilon_{i} c_{i}^{\dagger}c_{i} + 
\hspace{-1.5mm}\sum_{k\in\text{L,R}} 
\epsilon_{k} c_{k}^{\dagger}c_{k} + \sum_{k,i} ( V_{ki} 
c^{\dagger}_{k} c_{i} + \text{H.c.} ),\hspace{3mm} \\
\label{Hvib}
H_{\text{vib}} \hspace{-1.5mm} &=& \hspace{-1.5mm} \sum_{\alpha} 
\Omega_{\alpha} 
a_{\alpha}^{\dagger}a_{\alpha} + \sum_{\alpha,i} \lambda_{i\alpha} 
(a_{\alpha}+a_{\alpha}^{\dagger}) c_{i}^{\dagger}c_{i}. \hspace{3mm}
\end{eqnarray}
The electronic part of the Hamiltonian, $H_{\text{el}}$, 
includes electronic states with energies $\epsilon_{i}$
located at the molecular bridge (cf.\ Fig.\ \ref{scetchofmodelsystem}a). 
These states 
are coupled by interaction matrix elements $V_{ki}$ 
to electronic states with energies $\epsilon_{k}$ in the
leads. Thereby, the operators 
$c^{\dagger}_{i}$/$c_{i}$ ($c^{\dagger}_{k}$/$c_{k}$) represent 
the respective creation/annihilation operators for 
the states of the molecular bridge (the leads).
The vibrational degrees of freedom of the junction are described 
as harmonic oscillators 
with frequencies $\Omega_{\alpha}$, 
and corresponding creation/annihilation operators $a_{\alpha}^{\dagger}$/$a_{\alpha}$. 
Here, $\lambda_{i\alpha}$ denotes the electronic-vibrational 
(vibronic) coupling strength 
between vibrational mode $\alpha$ and the $i$th state of the bridge.

To study quantum interference and decoherence effects in molecular junctions, 
we employ a Non-Equilibrium Green's Function approach (NEGF) 
\cite{Galperin06,Hartle,Hartle09,Hartle2010,Hartle2010b}. 
NEGF theory allows to describe 
quasi-degenerate molecular levels, which is
crucial for the analysis of the respective quantum interference effects. 
The method is based on (self-consistent) second-order perturbation theory 
in the molecule-lead coupling, 
and treats electronic-vibrational coupling non-perturbatively. 
The details of our NEGF approach  
have been outlined previously \cite{Galperin06,Hartle,Hartle09,Hartle2010}. Briefly,  
the approach is based on the small polaron transformation \cite{Galperin06,Hartle,Hartle2010b}. 
The accordingly transformed Hamiltonian 
$\overline{H}$ 
comprises a polaron-shift of the electronic states of the molecular bridge, 
$\overline{\epsilon}_{i}=\epsilon_{i}-\sum_{\alpha}
(\lambda_{i\alpha}^{2}/\Omega_{\alpha})$,
a molecule-lead 
coupling term,
which is renormalized by the shift operators 
$X_{i}=\text{exp}(\sum_{\alpha}(\lambda_{i\alpha}/\Omega_{\alpha})
(a_{\alpha}-a_{\alpha}^{\dagger}))$, 
and Hubbard-like electron-electron interaction terms, 
${(\lambda_{i\alpha}\lambda_{j\alpha}/\Omega_{\alpha})c_{i}^{\dagger}
c_{i}c_{j}^{\dagger}c_{j}}$, 
but no direct electronic-vibrational coupling terms. 
The vibrationally induced electron-electron interactions are
treated using a non-perturbative approximate scheme \cite{Hartle09}. 
In the (anti-)adiabatic regime of a molecular junction, 
the electronic Green's function matrix $\textbf{G}$ 
can be described as 
$\textbf{G}_{ij}(\tau,\tau') 
 \approx \textbf{G}_{c,ij}(\tau,\tau') \langle 
\text{T}_{c} X_{i}(\tau)X^{\dagger}_{j}(\tau')\rangle$.  
The matrix $\textbf{G}_{c,ij}(\tau,\tau')=-i \langle 
\text{T}_{c} c_{i}(\tau) c_{j}^{\dagger}(\tau') \rangle$ 
is determined by the self-energy matrix 
$\mathbf{\Sigma}_{\text{L/R},ij}(\tau,\tau')=
\sum_{k\in\text{L/R}}V_{ki}^{*}V_{kj}g_{k}(\tau,\tau') \langle 
\text{T}_{c} X_{j}(\tau')X^{\dagger}_{i}(\tau)\rangle $, 
where $g_{k}$ denotes the free Green's function 
of lead state $k$ and 
$T_{c}$ is the time-ordering operator on the Keldysh contour. 
The correlation functions of the shift-operators $X_{i}$ 
are evaluated using a cumulant expansion 
in the dimensionless coupling parameters 
$\lambda_{i\alpha}/\Omega_{\alpha}$, 
which in turn requires the electronic Green's functions $\textbf{G}_{c,ij}$. 
Therefore, we employ a self-consistent solution scheme \cite{Galperin06,Hartle}. 
This scheme accounts for the molecule-lead coupling 
in second-order (self-consistent) perturbation theory. 
It is thus exact for vanishing vibronic coupling, 
and due to the small polaron transformation, also 
for vanishing molecule-lead coupling. 
The electrical current through the junction is given by 
$I=2e\int\left(\text{d}\epsilon/(2\pi)\right)\hspace{1mm} \text{tr}\lbrace
\mathbf{\Sigma_{\text{L}}^{<}}\textbf{G}_{c}^{>}-
\mathbf{\Sigma_{\text{L}}^{>}}\textbf{G}_{c}^{<} \rbrace$ \cite{Galperin06,Hartle,Hartle09}.

First, we consider a generic model system similar 
to the one depicted in Fig.\ \ref{scetchofmodelsystem}a with parameters that
are typical for single-molecule junctions. 
The model comprises two electronic states located 
at $\overline{\epsilon}_{1}=0.404$\,eV and $\overline{\epsilon}_{2}=0.4$\,eV 
above the 
Fermi-level $\epsilon_{\text{F}}=0$. 
State $1$ of our model system is coupled to the vibrational mode 
with coupling strength $\lambda_{1}=0.06$\,\text{eV}. 
For state $2$, we consider $\lambda_{2}=0$. This choice simplifies 
the analysis of the corresponding 
decoherence mechanism, which occurs in general for $\lambda_{1}\neq\lambda_{2}$. 
These states are coupled to 
the leads in the same way 
as shown in Fig.\ \ref{scetchofmodelsystem}a, \emph{i.e.}\ with 
$\nu_{\text{L},1/2}=\nu_{\text{R},1}=-\nu_{\text{R},1}=\nu=0.1$\,eV. 
The leads are modelled as semi-infinite tight-binding chains 
with level-width functions $\mathbf{\Gamma}_{\text{L/R},ij}(\epsilon)=(\nu_{\text{L/R,i}}
\nu_{\text{L/R,j}}/\gamma^{2})
\text{Re}\left[\sqrt{4\gamma^{2}-(\epsilon-\mu_{\text{L/R}})^{2}}\right]$ 
and $\gamma=3$\,eV. The bias voltage $\Phi$ is assumed to symmetrically 
drop at the contacts,  $\mu_{\text{L}}=-\mu_{\text{R}}=\Phi/2$.

\begin{figure}
\includegraphics[scale=\newscalex]{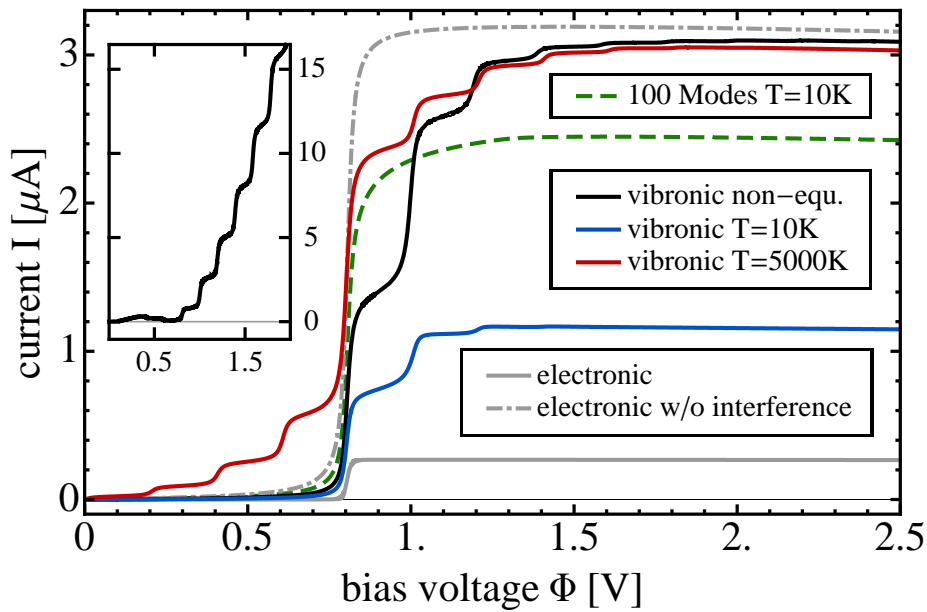}
\caption{\label{twostates} (Color online) Current-voltage characteristics 
for a generic model molecular junction that exhibits two destructively 
interfering electronic states (see Fig.\ \ref{scetchofmodelsystem}) 
coupled to a single vibrational mode. The inset shows the corresponding 
vibrational excitation characteristics, $\langle a^{\dagger}a \rangle$.}
\end{figure}

\begin{figure}
\includegraphics[scale=\newscalex]{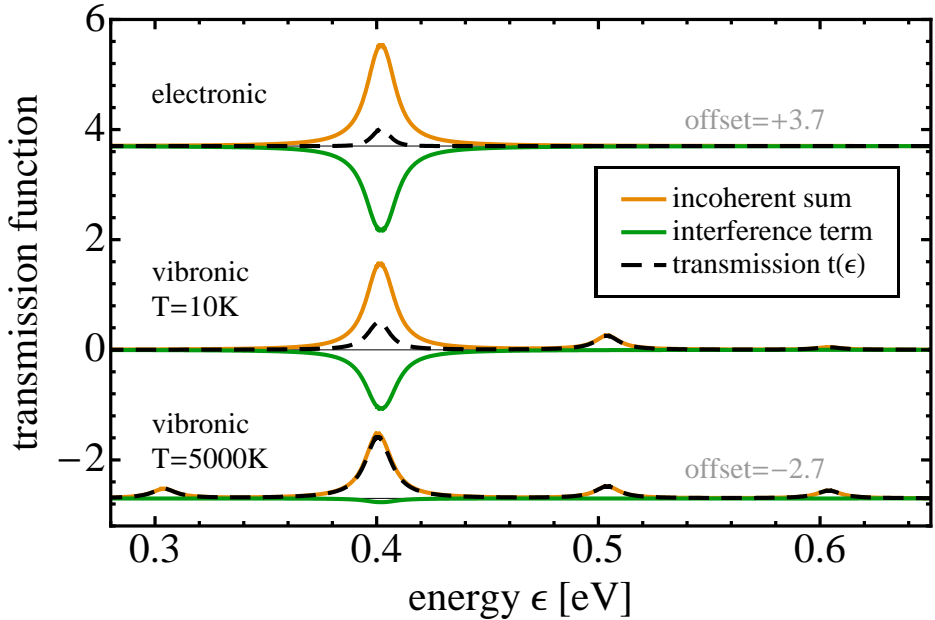}
\caption{\label{condes} (Color online) Transmission functions $t(\epsilon)$ 
(dashed black line) corresponding to the currents depicted by the solid  
gray, blue and red lines in Fig.\ \ref{twostates} at a high bias voltage. 
For clarity, the graphs for these three different cases are separated 
by offsets. In addition, the incoherent sum of transmission channels 
$t_{\text{sum}}(\epsilon)$ (orange lines) and the interference term 
$t_{\text{int}}(\epsilon)$ (green lines) are depicted 
(cf.\ Eqs.\ (\ref{condesparts})).}
\end{figure}

Current-voltage characteristics for this model system 
are shown in Fig.\ \ref{twostates}. 
Thereby, the solid gray line shows the electrical current obtained for static
nuclei ('electronic current'), \emph{i.e.}\ for 
the polaron-shifted electronic states with $\lambda_{1}=0$. 
The comparatively small value of the electronic current ($0.26$\,$\mu$A for
$\Phi > 2\overline{\epsilon}_{1/2}$) is a result of destructive interference effects 
in this molecular junction, which arise due to the symmetry of the 
molecule-lead coupling and the quasidegeneracy of the 
two states ($\overline{\epsilon}_{1}-\overline{\epsilon}_{2}\lesssim\Gamma$). 
If the partial currents corresponding to the different
pathways through the molecular junction are
added incoherently (dashed gray line), a much larger value  ($3.2$\,$\mu$A for 
$\Phi > 2\overline{\epsilon}_{1/2}$) is
obtained. This demonstrates the presence of 
destructive quantum interference, 
which reduces the current through this junction 
by more than an order of magnitude.

If we take into account electronic-vibrational coupling, $\lambda_{1}=0.06$\,eV, 
vibrational excitation and deexcitation processes 
(cf.\ Fig.\ \ref{scetchofmodelsystem}b and \ref{scetchofmodelsystem}c) 
contribute to the electron transport through the junction 
\cite{Hartle,Hartle09,Hartle,Hartle2010,Hartle2010b}. 
These inelastic processes result 
in an highly excited state of the vibrational mode for $\Phi\gtrsim2\epsilon_{1}$ 
(cf.\ the vibrational excitation characteristics shown in the inset of 
Fig.\ \ref{twostates}). 
The respective current-voltage characteristics, 
which is depicted by the solid black line in Fig.\ \ref{twostates}, 
exhibits several steps at $\Phi=2(\overline{\epsilon}_{1}+n\Omega)$, 
$n\in\mathbb{N}$, corresponding to vibrationally-assisted electron 
transport processes \cite{Hartle,Hartle09,Hartle2010,Hartle2010b}. 
Intriguingly, the current that is obtained including electronic-vibrational 
coupling (black line),  
is much larger than the electronic current (gray line) and approaches the 
value obtained by incoherent summation of the 
different electronic pathways (dashed gray line). 
Electronic-vibrational coupling thus results in a complete quenching of 
destructive interference effects in this model molecular junction, 
or equivalently, leads to strong decoherence. 
While the results discussed so far have been obtained using the
full current-induced nonequilibrium vibrational distribution, the basic 
mechanism of vibronic decoherence can also be described 
within the often employed  approximation,  where the 
state of the  vibrational mode is restricted to its thermal equilibrium state 
after each electron transmission event. 
The solid blue and red lines in Fig.\ \ref{twostates} 
show results employing this simpler approach obtained for temperatures of 
$10$\,K and $5000$\,K \cite{noteOn5000K}, respectively.

To analyze these effects in more detail we 
employ a thermal equilibrium state for the vibration 
and consider bias voltages $\Phi\gg\epsilon_{1/2}$, for which 
the current can be expressed in a form similar to Landauer theory, 
$I\approx2e\int_{-\infty}^{\infty}(\text{d}\epsilon/2\pi)\,t(\epsilon)$, 
with the transmission function 
$t(\epsilon)\equiv i\,\text{tr}\lbrace \mathbf{\Gamma}_{\text{L}}  
\textbf{G}^{>}   \rbrace$. 
This transmission function $t(\epsilon)$ can be split into an incoherent 
sum of transmission channels,  
$t_{\text{sum}}(\epsilon)=i\sum_{i} \tilde{\textbf{G}}\hspace{0mm}^{>}_{ii}
\mathbf{\Gamma}_{\text{L},ii}$, where $\tilde{\textbf{G}}\hspace{0mm}^{>}_{ii}$ is evaluated disregarding 
the off-diagonal elements of $\mathbf{\Sigma}_{\text{L/R}}$, and an interference term,  
$t_{\text{int}}(\epsilon)=t(\epsilon)-t_{\text{sum}}(\epsilon)$. 
The interference term explicitly shows the interference effects between 
individual transmission channels 
encoded in the off-diagonal elements of $\mathbf{\Sigma}_{\text{L/R}}$. 
For our specific model system, in the wide-band approximation 
with $\Gamma=2\nu^{2}/\gamma$, 
we obtain  
\begin{eqnarray}
\label{condesparts}
 t_{\text{sum}}(\epsilon) &=& (\Gamma/\left\vert 
\epsilon-\overline{\epsilon}_{2} + i \Gamma\right\vert)^{2} \\
&& \hspace{-0.2cm} +  A  \sum_{l=-\infty}^{\infty} I_{l}(x) 
\text{e}^{\beta l\Omega/2} (\Gamma/\left\vert \epsilon-\overline{\epsilon}_{1} - 
l \Omega + i \Gamma \right\vert)^{2}, \nonumber\\
 t_{\text{int}}(\epsilon) &=& -2 A  \Gamma^{2} \text{Re}\left[ ( \epsilon-
\overline{\epsilon}_{1} + i \Gamma )^{-1}( \epsilon-\overline{\epsilon}_{2} - 
i \Gamma )^{-1} \right]. \nonumber
\end{eqnarray}
Thereby, the prefactor $A=\text{e}^{-(\lambda_{1}^{2}/\Omega^{2})(2N_{\text{vib}}+1)}$ 
is determined by
the average vibrational excitation 
$N_{\text{vib}}=(\text{e}^{\beta\Omega}-1)^{-1}$ and the inverse temperature 
$\beta=(k_{\text{B}}T)^{-1}$, while  
$I_{l}(x)=I_{l}(2(\lambda_{1}^{2}/\Omega^{2})\sqrt{N_{\text{vib}}(N_{\text{vib}}+1)})$ 
denotes the $l$th modified Bessel function of the first kind.

Fig.\ \ref{condes} shows the transmission functions (dashed black lines)
corresponding to the currents  
represented by the gray, 
the solid blue and red line in Fig.\ \ref{twostates}. 
In addition to the transmission functions, 
the incoherent sum of transmission channels (orange line) and the interference 
terms (green line) are depicted. 
Two important observations can be made. 
First, for non-vanishing vibronic coupling,  
the transmission peak associated with state $1$ is 
fragmented into vibrational side-peaks. 
These side-peaks, which are associated with vibronic tunneling processes 
(Fig.\ \ref{scetchofmodelsystem}b and \ref{scetchofmodelsystem}c), 
appear in the incoherent sum of transmission channels, 
but not in the interference term. 
Thus, these side-peaks can 
directly contribute to the total transmission probability. 
Second, the prefactor 
$A=\text{e}^{-(\lambda_{1}^{2}/\Omega^{2})(2N_{\text{vib}}+1)}$ 
reduces the main peak at $\epsilon\approx\overline{\epsilon}_{1/2}$ 
in both parts of the transmission function. 
While for $N_{\text{vib}}\rightarrow\infty$ the elastic peak in the 
incoherent sum of the transmission channels is thus halved, 
the peak in the interference term completely vanishes for a high level 
of vibrational excitation.  
As a result, quantum interference effects are greatly reduced, 
if the vibration acquires a highly excited state, 
in particular in the resonant tunneling regime 
(cf.\ the inset of Fig.\ \ref{twostates}).  

In the model considered so far, the decoherence mechanism is based 
on the coupling of the electronic states of the molecular junction 
to its internal vibrational degrees of freedom. 
In many cases \cite{Segal2000,Lehmann04}, 
however, 
decoherence (or dephasing) is a result of 
external or environmental degrees of freedom. 
Such a mechanism can be described by a different model, 
where the electronic states are coupled to 
a bath of harmonic oscillators, mimicking, e.g., the coupling 
to the phonons in the leads \cite{Seidemann10} 
or a surrounding solvent in an electrochemically 
gated molecular junction \cite{Tao2006,Seidemann10}. 
The current-voltage characteristics obtained for such a model is shown in 
in Fig.\ \ref{twostates} by the dashed green line. Thereby, the overall 
vibronic coupling strength (as specified by the reorganization energy, $0.36$\,meV) 
is the same as for the single-mode model considered above but has been 
equally distributed over $100$ vibrational modes, which frequencies range 
from $1$\,meV to $250$\,meV with an equidistant spacing of $2.5$\,meV. 
The comparison to the corresponding single-mode case  (solid blue line) 
shows that the coupling to a multitude of vibrational modes induces even 
stronger decoherence effects. 
Within the analysis of interference discussed above, 
this mechanism is encoded in the prefactor $A$ of Eqs.\ (\ref{condesparts}), 
which for the present model reads  
$A=\text{e}^{-\sum_{\alpha}(\lambda_{1\alpha}^{2}/\Omega_{\alpha}^{2})
(2N_{\text{vib},\alpha}+1)}$. Due to the low-frequency modes, it 
is considerably smaller than for 
the single-mode model.

\begin{figure}
\begin{flushleft}
\text{}\\[0.05cm]
\text{a)\hspace{\shifta}HOMO \hspace{\shiftb}  HOMO-1 \hspace{\shiftc} HOMO-4} \\
\end{flushleft}\vspace{-0.4cm}
\includegraphics[scale=\newscaley]{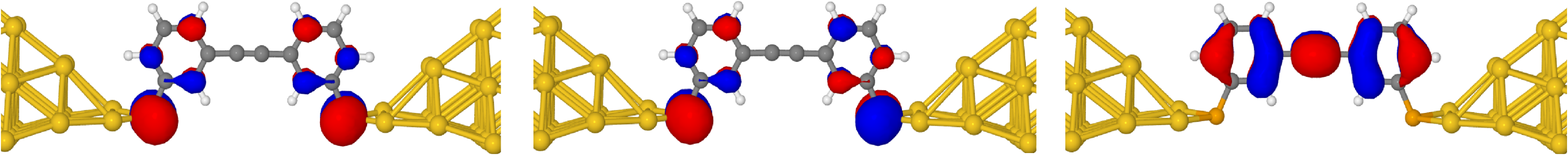}\\
\begin{flushleft}
b) 
\end{flushleft}
\vspace{-1.35cm} \includegraphics[scale=\newscalex]{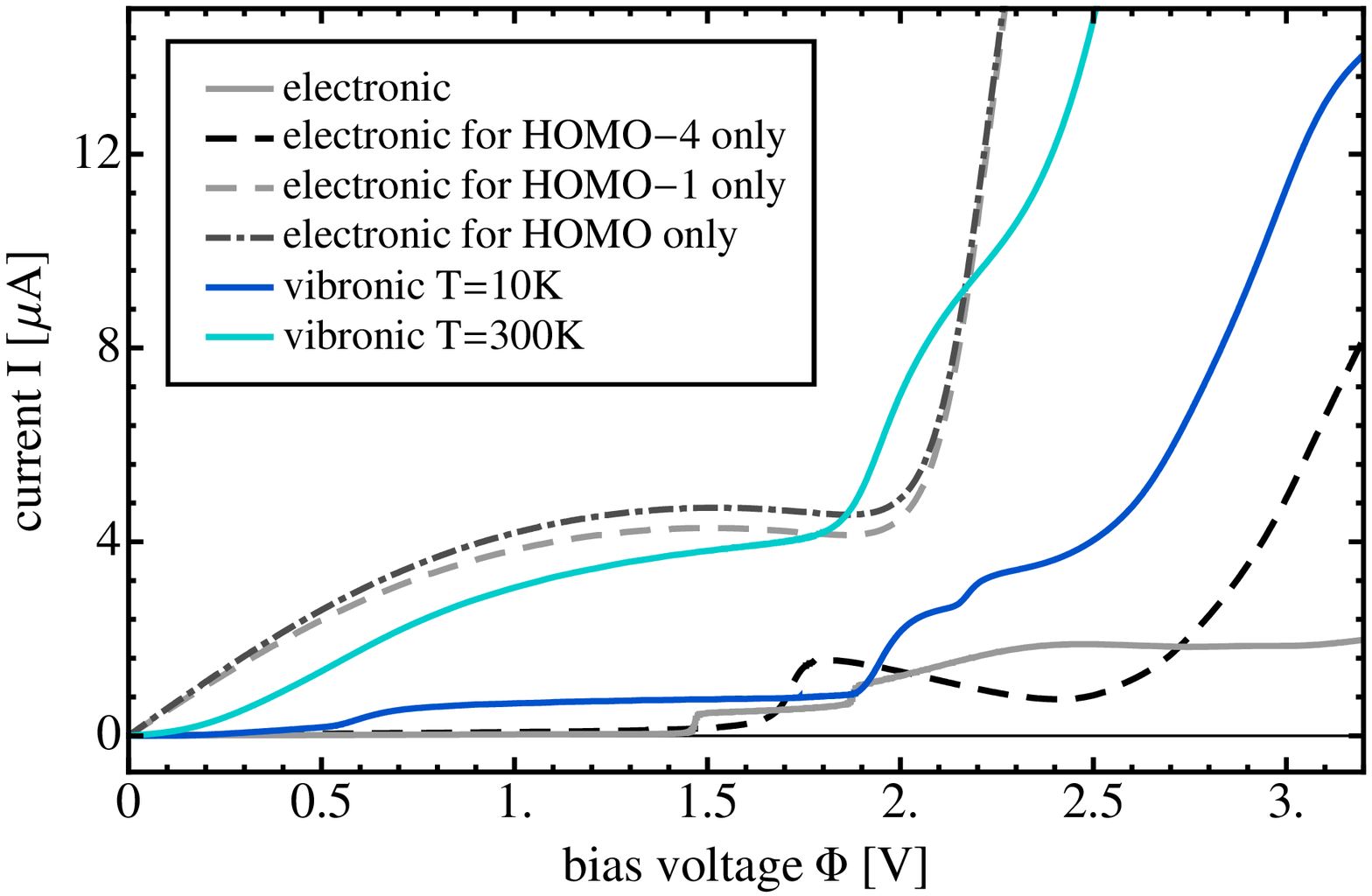}
\caption{\label{ortho} (Color online) 
Electron Transport in an \emph{o}-biphenylacetylenedithiolate molecular junction. 
a): relevant molecular orbitals, where the blue and the red color 
represent different signs. 
b): current-voltage characteristics for 
different scenarios as described in the legend. 
}
\end{figure}

The results discussed so far for a generic model system show that
electronic-vibrational coupling  leads to decoherence in single-molecule
junctions and elucidates the underlying mechanisms.
As a specific example for a single-molecule
junction that exhibits quantum interference and decoherence,  
we consider a  \emph{o}-biphenylacetylenedithiolate molecule bound 
to gold electrodes (see Fig.\ \ref{ortho}). 
The model parameters of this junction were determined by
first-principles electronic structure calculations, where a 
detailed description of our methodology 
can be found in Ref. \cite{Benesch08}. 
The junction exhibits 36 active vibrational modes 
coupled to a multitude of closely-lying electronic states 
that interfere with each other. 
Test calculations show that for bias voltages within a range
of $|\Phi|\lesssim3$\,V the three electronic states
corresponding to the
molecular orbitals depicted in Fig.\ \ref{ortho}a) are sufficient to describe the 
electronic current in this junction. 
The comparison of the electronic current obtained within this three-state
model (solid gray line)
with the significantly larger current that is obtained for
the transport through the individual states 
(dashed lines) demonstrates the presence of pronounced
interference effects among the three pathways in this junction. 
Analysis shows that the destructive interference of the 
highest  occupied molecular orbital (HOMO) and 
HOMO-1 are due to their quasidegeneracy and the fact that
their wavefunctions are symmetric and antisymmetric combinations,
respectively. In addition, the electronic state corresponding to 
HOMO-4 exhibits destructive interference with 
the other two orbitals, due to the broadening of the molecular levels.
If we account for electronic-vibrational coupling, 
the current-voltage characteristics represented by the solid blue 
and turquoise line in Fig.\ \ref{ortho} are obtained, where all 36 vibrational 
degrees of freedom  are evaluated in thermal equilibrium at $10$\,K and $300$\,K, respectively. 
The associated current-voltage characteristics are 
significantly larger than the electronic current. 
Moreover, for an increasing temperature of the vibrational modes, 
the current through the junction approaches
the significantly higher values obtained for the individual channels, given by the dashed lines. 
As in the generic model system, vibronic interactions thus result in 
a strong decoherence of interfering 
electronic states, which is greatly enhanced by the temperature of the modes 
or equivalently their level of vibrational excitation.

In summary, single-molecule junctions often exhibit multiple  
closely-lying electronic states that 
provide interfering pathways for the tunneling of electrons. 
Our analysis, obtained for both a generic model and a specific example, show that 
electronic-vibrational coupling leads to decoherence and quenching of
these electronic quantum interference effects. 
As a result, the electrical current in molecular junctions can be
significantly  larger than without electronic-vibrational coupling.  
This effect is particularly pronounced if the junction is vibrationally
highly excited, e.g.\ due to current-induced
nonequilibrium effects in the resonant transport regime.
It should also be emphasized that this decoherence mechanism is an
intrinsic  property of a single-molecule junction.

We gratefully acknowledge fruitful discussions with S.\ Ballmann, 
B.\ Kubala and P.\ B.\ Coto, 
and generous allocation of computing time by the Leibniz Rechenzentrum 
M\"unchen (LRZ). 
The work at the Friedrich-Alexander 
Universit\"at Erlangen-N\"urnberg was carried out in the
framework of the Cluster of Excellence "Engineering of Advanced Materials".


\begin{thebibliography}{27}
\bibitem[{not()}]{noteOn5000K}
\bibinfo{note}{At $5000$\,K the average thermal excitation of the vibrational
  mode is comparable to the one obtained by the full nonequilibrium calculation
  (cf.\ the inset of Fig.\ \ref{twostates}).}
\end{thebibliography}
\end{document}